\documentclass[10pt,twocolumn,letterpaper]{article}

\usepackage{iccv}              

\usepackage{graphicx} 
\usepackage{array,booktabs} 
\usepackage{amsfonts, amsmath, amssymb}
\usepackage{multirow}
\usepackage{subcaption}
\usepackage{fancyhdr}
\usepackage[T1]{fontenc}
\usepackage[dvipsnames]{xcolor}
\usepackage[inline]{enumitem}
\usepackage{ragged2e}
\usepackage{pseudo}
\usepackage{mdframed}
\usepackage{tikz}
\usepackage{tabularx}
\usepackage{mathtools}
\usepackage{authoraftertitle}
\usepackage[final]{changes}
\setanonymousname{robin}
\newcommand{\RsevenrAT}{\noindent\textcolor{orange}{\textbf{7rAT}}}
\newcommand{\JXeightR}{\noindent\textcolor{magenta}{\textbf{JX8R}}}

\newcommand{\abssq}[1]{|#1|^2}

\newcommand{\defphase}[1]{\mathcal{#1}}

\newcommand{\deffourier}[1]{\defphase{F}\left\{{#1}\right\}}
\newcommand{\defconvolve}{\circledast}

\newcommand{\twodPixels}{\left(x, y\right)}
\newcommand{\threedPixels}{\left(x, y, t\right)}
\newcommand{\uvtPixels}{\left(u, v, t\right)}

\newcommand{\capturedImage}{I}
\newcommand{\objectImage}{O}

\newcommand{\atmospherePhase}{\defphase{A}}
\newcommand{\dmPhase}{\defphase{D}}

\newcommand{\systemPSF}{\defphase{P}} 
\newcommand{\idealPSF}{\defphase{P}_{\textsc{\tiny ideal}}}
\newcommand{\effectivePSF}{\defphase{P}_{\textsc{\tiny eff}}} 
\newcommand{\earthPSF}{\defphase{P}_{\textsc{\tiny TEL}}}
\newcommand{\correctedPSF}{\defphase{P}_{\textsc{\tiny corr}}}
\newcommand{\shiftPSF}{\defphase{S}} 
\newcommand{\zernPSF}{\defphase{Z}}

\newcommand{\funcParamters}{\theta}
\newcommand{\funcUpsample}[1][x]{f_\Uparrow\left(#1~|~\funcParamters\right)}
\newcommand{\funcDownsample}[1][x]{f_\Downarrow\left(#1\right)}

\newcommand{\imageNoise}{\eta}

\renewcommand{\arraystretch}{1.5}

\definecolor{iccvblue}{rgb}{0.21,0.49,0.74}
\usepackage[pagebackref,breaklinks,colorlinks,allcolors=iccvblue]{hyperref}

\def\confName{ICCV}
\def\confYear{2025}

\title{Super Resolved Imaging with Adaptive Optics}

\usepackage{float}
\usepackage[percent]{overpic}

\newcommand{\edit}[1]{\textcolor{black}{#1}} 

\newcommand\blfootnote[1]{%
  \begingroup
  \renewcommand\thefootnote{}\footnote{#1}%
  \addtocounter{footnote}{-1}%
  \endgroup
}

\author{
  Robin~Swanson\textsuperscript{1,\,2}\quad
  Esther~Y.~H.~Lin\textsuperscript{1}\quad
  Masen~Lamb\textsuperscript{3,\,4}\quad 
  Suresh~Sivanandam\textsuperscript{1,\,2}\quad
  Kiriakos~N.~Kutulakos\textsuperscript{1}\\[0.5em]
  \textsuperscript{1}\,University of Toronto\quad
  \textsuperscript{2}\,Dunlap Institute for Astronomy \& Astrophysics\\[-0.4em]
  \textsuperscript{3}\,International Gemini Observatory\quad
  \textsuperscript{4}\,University of Victoria
}

\begin{document}

\maketitle
\begin{abstract}
Astronomical telescopes suffer from a tradeoff between field of view (FoV) and image resolution: increasing the FoV leads to an optical field that is under-sampled by the science camera. This work presents a novel computational imaging approach to overcome this tradeoff by leveraging the existing \emph{adaptive optics (AO)} systems in modern ground-based telescopes. Our key idea is to use the AO system’s deformable mirror to apply a series of learned, precisely controlled distortions to the optical wavefront, producing a sequence of images that exhibit distinct, high-frequency, sub-pixel shifts. These images can then be jointly upsampled to yield the final super-resolved image. Crucially, we show this can be done while simultaneously maintaining the core AO operation---correcting for the unknown and rapidly changing wavefront distortions caused by Earth's atmosphere. To achieve this, we incorporate end-to-end optimization of both the induced mirror distortions and the upsampling algorithm, such that telescope-specific optics and temporal statistics of atmospheric wavefront distortions are accounted for. Our experimental results with a hardware prototype, as well as simulations, demonstrate significant SNR improvements of up to 12\,dB over non-AO super-resolution baselines, using only existing telescope optics and no hardware modifications. Moreover, by using a precise bench-top replica of a complete telescope and AO system, we show that our methodology can be readily transferred to an operational telescope.
\end{abstract}

\blfootnote{Corresponding author: \url{robin@cs.toronto.edu}}
\blfootnote{Project website: \url{www.cs.toronto.edu/~robin/aosr}}    
\vspace{-.8cm}
\section{Introduction}
\label{sec:intro}
\vspace{-0.3cm}
Super-resolution \edit{(SR)} techniques have been integral to astronomical imaging due to inherent telescope design constraints~\cite{fruchter2002drizzle}. \replaced{Telescopes can have many science instruments and in certain cases optical resolution is not their priority~\cite{zhu2019revisiting}. A large \edit{field of view (FoV)}, for example, necessitates undersampling to capture a wider view of the night sky. Another common case in photon-starved conditions is to increase \edit{signal-to-noise ratio} by focusing the light onto fewer, larger, pixels, thereby incurring less read noise~\cite{mellier2024euclid}. In these, and many other, cases although the telescope is capable of providing a diffraction-limited signal, the science instrument can be undersampled. Thus, the use of \edit{SR} methods in astronomical imaging has a long history in both satellite and terrestrial telescopes~\cite{lauer1999combining}.}{Super-resolution techniques have been integral to astronomical imaging due to inherent telescope design constraints~\cite{fruchter2002drizzle}. Certain parameters, such as the measured wavelength of light or field of view, may be prioritized in the design of a telescope or science instrument~\cite{zhu2019revisiting}. Often, these considerations necessarily limit the spatial resolution on the science camera due to physical or signal-to-noise requirements~\cite{mellier2024euclid}. Thus, the use of super-resolution methods in astronomical imaging has a long history in both satellite and terrestrial telescopes~\cite{lauer1999combining}.}

Unlike telescopes stationed in orbit, ground telescopes face an additional challenge: the Earth’s atmosphere~\cite{beckers1993adaptive}. Our atmosphere is a source of dynamic optical aberrations which introduce additional difficulties regardless of a telescope’s location, size, or design~\cite{tyson2000adaptive}. As the atmosphere evolves, small changes in index of refraction occur across the sky. This causes incoming light to distort as it transitions from outer space and through our telescope, blurring the images. Therefore, traditional multi-image \edit{SR} methods which combine multiple sharp images with small geometric distortions—whether from telescope movements~\cite{fruchter2002drizzle} or Earth’s rotation~\cite{bauer2011super} are not well suited for these conditions.

We present Super Resolved Imaging with Adaptive Optics, an end-to-end \edit{SR} method for terrestrial telescopes \edit{that} leverages existing adaptive optics (AO) systems—designed to correct for atmospheric distortions—to simultaneously correct for aberrations and achieve \edit{SR}. The crux of our approach is to take advantage of the existing \edit{AO} system in ground telescopes, which applies a phase mask in the telescope’s Fourier domain to counteract atmospheric aberrations. We use the AO system by optimizing and injecting additional phase displacements that produce sub-pixel shifts in the image plane. After capturing several images, these shifts serve as inputs to \edit{an} end-to-end trained, multi-image \edit{SR} (MISR) method. These additional phases are optimized to minimally alter the telescope’s point spread function (PSF), thus subtly modifying each exposure to best facilitate MISR. The phase displacements can be co-optimized with \edit{deep learning–based} (or any differentiable) MISR to enhance overall performance. 
\edit{This optimization is performed entirely offline, requiring no modifications to the telescope or use of on-sky time.}

In summary, our contributions include:
\begin{itemize}[leftmargin=.25in]
\item A novel MISR method for telescopes that leverages existing hardware, requires no modifications, and does not impact the science payload.
\item Demonstration that combining AO with MISR outperforms either technique alone, achieving performance greater than the sum of its parts.
\item Showing that phase displacements and image reconstruction can be co‑optimized end‑to‑end offline using simple telescope telemetry as a proxy for real systems, enabling unobtrusive deployment where telescope time is extremely costly~\cite{kulkarni2016instruments}.
\item A real‑hardware setup demonstrating the feasibility of integrating our method into existing telescope systems.
\end{itemize}
\vspace{-1em}
\section{Principles of Adaptive Optics}
\label{sec:ao_primer}
\vspace{-0.3cm}
The Earth's atmosphere fundamentally \edit{limits} the resolving power of any telescope {by introducing optical aberrations}---one of the reasons why {space-based} telescopes such as the Hubble Space Telescope have outperformed many larger ground-based observatories. 
This results in the final telescope PSF being several factors worse than the diffraction limit—the best achievable resolution. As such, AO systems are now widely used in modern ground-based optical and near-infrared telescopes to reverse these effects.

\vspace{-1em}
\subsection{Components of an AO system}
\vspace{-0.2cm}

\begin{figure}[!ht]
    \centering
    \includegraphics[width=0.475\textwidth]{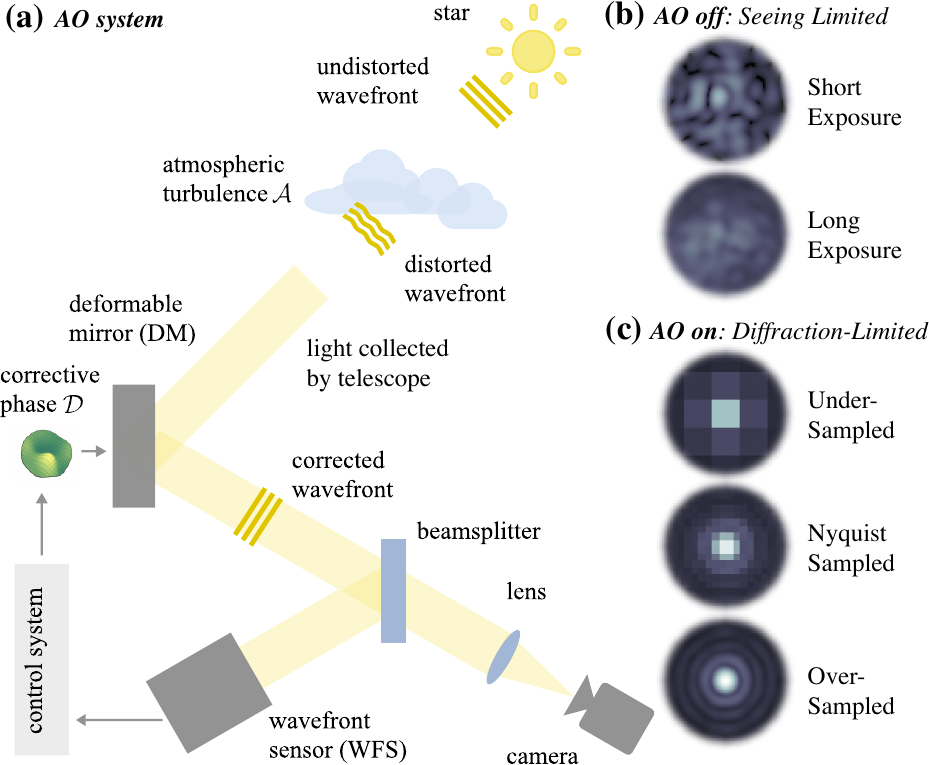}
    \caption[{Illustration of a prototypical AO system and the effect of AO on telescope PSF. }]{
    Starlight is aberrated by Earth's atmosphere \(\atmospherePhase\), blurring the telescope PSF.
    An AO system (\edit{\textbf{a}}) measures the atmosphere with a wavefront sensor and applies a corrective phase, \(\dmPhase\), on its deformable mirror to cancel it out.
    When AO is off, resolution is directly limited by the seeing conditions (i.e., turbulence) at the observatory for both short and long exposures (\edit{\textbf{b}}). 
    When AO is enabled the effects of the atmosphere are removed. 
    If sampled at or above the Nyquist rate, the resolution is limited only by the mirror diameter and its PSF forms a \edit{diffraction-limited} Airy disk (\edit{\textbf{c}}). 
    Our method enables \edit{under-sampled} instruments with AO to recover spatial resolution.} 
    \label{fig:ao_schematic}
\end{figure}

AO systems correct atmospheric turbulence by introducing a phase equal \edit{in magnitude} but opposite in sign to that of the atmosphere. A prototypical AO setup (\cref{fig:ao_schematic}) consists of a deformable mirror (DM), wavefront sensor (WFS), and control system. The DM, {placed on the Fourier plane of the telescope,} takes on the opposite {phase} of the atmosphere, resulting in corrected wavefronts of light which are then passed on to the science camera and back to the WFS. Because the atmosphere evolves over time, the WFS must continuously measure the residual, uncorrected wavefront, and send updated commands to the DM. This loop, implemented via the control system, typically operates on the order of 1-2 kHz for modern AO systems.

The most common types of WFSs are based on Shack-Hartmann~\cite{platt2001history} and pyramid~\cite{ragazzoni2002pyramid} designs. To operate they require a ``guide-star''\edit{---a distant light source which can be approximated as a point.} \replaced{Usually a bright star \edit{near the science object is used}, but laser beacons can also be used to create an artificial star image in the upper atmosphere. } 
\edit{The guide star samples a similar turbulence path as the science image without reducing its signal-to-noise ratio (SNR).}
By comparing the measured signal to a reference, we can calculate how the atmosphere has affected the wavefront across the telescope pupil. 
\edit{Pre-calibrated responses are then used to update the DM shape and negate the distortions~\cite{tyson2000adaptive}.}

\edit{The DM can be constructed in several ways. Many of today’s telescopes use a continuous design with a single flexible reflective membrane supported by a grid of discrete actuators~\cite{ealey1990continuous}. These actuators can be individually pushed or pulled to collectively generate complex phase patterns. In contrast, many upcoming AO systems use grids of individual mirrors, enabling much larger DMs and corrections for even higher spatial frequencies~\cite{gilmozzi2007european}.}

\edit{Once converged, the effects of the AO loop, shown in \cref{fig:ao_schematic}, are visible and have enabled many recent breakthroughs in modern astronomy.} 
Direct imaging of exo-planets~\cite{marois2008direct} and the first direct evidence of a super-massive black hole within our galaxy~\cite{ghez1998high} would not have been possible without the increased resolution and contrast provided by AO. For additional details on how AO has been recently used in both astronomy and microscopy see~\cite{hampson2021adaptive}.

\vspace{-1em}
\subsection{The AO Image Formation Model}
\vspace{-0.2cm}
Suppose we are observing a distant, constant object (e.g., a natural guide star) over an exposure time \(T\) using a telescope without AO. The recorded image \(I(x,y)\) in pixel coordinates \((x,y)\) is modeled as the convolution of the object \(O(x,y)\) with an atmosphere-degraded PSF \(\earthPSF\). Here, \(\earthPSF\) corresponds to a diffraction-limited PSF \(\idealPSF\) (i.e., an Airy Disk~\cite{tyson2022principles}) degraded by the time-varying turbulence of the Earth's atmosphere. In Fourier space, this turbulence can be represented by a phase distribution \(\atmospherePhase\uvtPixels\) with unit amplitude where \(\left(u, v\right)\) are its spatial frequency coordinates. By the convolution theorem, this can be described in Fourier space with a point-wise multiplication:
\begin{equation}
    \deffourier{\earthPSF}\uvtPixels = e^{\edit{j}\atmospherePhase\uvtPixels}\deffourier{\idealPSF}\uvtPixels,
\end{equation}
where $\deffourier{\cdot}$ denotes the Fourier Transform operator.  While the AO is in operation, a DM continuously applies an additional phase distortion $\dmPhase\uvtPixels$ in the Fourier plane of the telescope to counteract the effects of the atmosphere. The resulting corrected PSF, \(\correctedPSF\), can be described in the Fourier domain as 
\begin{equation}
\deffourier{\correctedPSF}\uvtPixels = e^{\edit{j}\dmPhase\uvtPixels} e^{\edit{j}\atmospherePhase\uvtPixels}\deffourier{\idealPSF}\uvtPixels.
\end{equation}
{In an ideal scenario, the atmospheric phase $\atmospherePhase$ and the deformable mirror phase $\dmPhase$ would perfectly cancel at every point and time; i.e., $\correctedPSF\approx\idealPSF$.}

{In practice the DM is limited by fitting errors and the finite range of its actuators which restricts the range of spatial frequencies it can correct. Additionally, the effects of non-circular mirror geometries~\cite{harvey1995diffraction} (e.g., hexagonal apertures) and mirror support structures cause the DM to deviate from ideal performance. Consequently, the final effective PSF of the telescope never truly reaches \(\idealPSF\). We model these effects on \(\effectivePSF\) in the Fourier domain as}
\begin{equation}
    \deffourier{\effectivePSF}\uvtPixels = e^{\edit{j}\delta\uvtPixels} \deffourier{\correctedPSF}\uvtPixels,
\end{equation}
where \(\delta\uvtPixels\) encapsulates the hardware and optical constraints described above. 

\added{Note that these contributions are predominantly static} and can \deleted{reasonably} be approximated as \(\delta\uvtPixels \approx \delta\left(u, v\right)\). \added[comment={\RsevenrAT: \textit{static $\delta(u,v)$ approximation [...] neglects temporal variations in hardware.}}]{For a well-performing, diffraction-limited AO system, variations will be short because the AO re-converges quickly. Therefore, over long exposures, they average out over time.} 
Finally, we consider incoherent image formation, as astronomical sources emit light independently with random phases. Our image formation becomes
\begin{equation}
    \capturedImage\twodPixels = \int_0^T  \abssq{\objectImage\twodPixels} \defconvolve \abssq{\effectivePSF\threedPixels}  \, \mathrm{d}t + \imageNoise
    \label{eq:image_formation_model}
\end{equation}
where $\defconvolve$ denotes the convolution operator, $T$ is the total exposure time, and $\imageNoise$ is cumulative noise. 

\vspace{-1em}
\section{Previous Work}
\label{sec:previous_work}
\vspace{-0.3cm}
\noindent\textbf{Image Super-Resolution (SR).}
SR techniques aim to reconstruct a high-resolution (HR) image from one or more observed low-resolution (LR) images ~\cite{yue2016image}. 
However, the problem is ill-posed in nature, and many plausible solutions exist~\cite{maral2022single}. 
SR methods are distinguished between single-image (SISR) and multi-image (MISR) super-resolution depending on the number of images used. 
Deep-learning SR has emerged as the state-of-the-art for both SISR and MISR tasks~\cite{maral2022single, yang2019deep, anwar2020deep, tian2011survey, maral2022single}, thanks to the growth of   convolutional networks and large datasets~\cite{dong2014learning, ledig2017photorealisticsingleimagesuperresolution}. However, deep SR networks thrive on natural images but struggle with out‑of‑distribution data due to limited training data~\cite{anwar2020deep}. In such cases, incorporating priors based on sensor, object, and acquisition conditions can improve performance~\cite{anwar2020deep}. Astronomical imaging is a regime with unique conditions, including extreme low-light, non-static aberrations, and \edit{image features unlike those typically seen in terrestrial images}. Hence, additional modifications to existing deep-learning SR techniques are required.

\noindent\textbf{SR in Astronomical Imaging:}
\texttt{Drizzle}—a SR technique commonly used in astronomical imaging and astrophotography—aligns and linearly blends multiple deliberately dithered frames to produce a super‑resolved image~\cite{fruchter2002drizzle}.
The technique has been widely adopted, including by the Hubble Space Telescope, and remains in use today~\cite{gonzaga2012drizzlepac}.
However, the success of drizzling and its contemporaries hinges on highly accurate image registration and precise knowledge of the offsets~\cite{borncamp2015ACSWFCRG, Avila2012AstroDrizzleAI}, a challenging task when the atmosphere adds time-varying aberrations and distortions~\cite{fruchter2011new}. In contrast, our method leverages the telescope's built-in AO system to inject optimized phase-domain ``jitters'' \replaced[comment={\RsevenrAT: claim of eliminating registration needs validation. there is no quantitative comparison with state-of-the-art drizzling techniques [...]; \RsevenrAT: superresolution and wavefront shaping?; \JXeightR,\RsevenrAT: Missing comparisons [to] Swanson et al. }]{and does not use any type of image registration. SR techniques have also been leveraged to improve the resolution and quality of WFS measurements~\cite{oberti2022super}. 
Our method, building on our previous work~\cite{swanson2022adaptive}, is agnostic to the underlying WFS type and can remain applicable as WFS technology advances.}{, eliminating the need for accurate registrations.}

\noindent\textbf{SR in Computational Imaging:}
Camera system design often incorporates super-resolution. One early example jointly optimized both the optics and SISR algorithms \replaced[comment={Reviewer comment: this was not a DOE}]{for simple, refractive lenses}{to improve resolutions in diffraction optical elements}~\cite{sitzmann2018end}.
Burst photography from handheld cameras---where small distortions are induced by hand shake---are extremely successful at recovering high-resolution information~\cite{wronski2019handheld, bhat2021deep} even in low-light conditions~\cite{liba2019handheld}. 
Deep learning methods have also been explored in the scientific imaging community for both microscopy~\cite{fang2021deep} and satellite imaging~\cite{nguyen2021self}.
MISR has also long played a role in classic camera development; many digital cameras are able to programmatically shift the imaging sensor between subsequent images which can then be combined into one higher resolution image (e.g., Sony’s Pixel Shift Multi Shooting~\cite{sonyPixelShift}). This work draws inspiration from previous works by incorporating end-to-end optimization of telescope mirror distortions with SR algorithms.

\noindent\textbf{Computational Wavefront Sensing and Control:}
CNNs have been used to improve WFS reconstruction~\cite{paine2018machine, dubose2020intensity}, as well as to predict atmospheric conditions to reduce servo-lag in the system~\cite{swanson2021closed, landman2020self}. Similar techniques have shown to be effective in microscopy AO systems as well~\cite{cumming2020direct, zhang2021deep}. More general-purpose methods have been proposed such as ~\cite{wang2018megapixel} which used coded wavefront sensors and spatial light modulators (SLMs) to achieve much higher order spatial frequency correction than is capable from typical AO systems in astronomy or microscopy. However, SLMs, unlike DMs, may not be suitable for correcting rapidly changing atmospheric conditions due to their slower speeds~\cite{chan2022holocurtains}.

\noindent\textbf{Optical Phase and PSF Optimization:}
End-to-end optimization of optical systems and their image reconstruction algorithms have been used in thin lens design~\cite{tseng2021neural}, holography~\cite{peng2020neural, chakravarthula2020learned}, and active illumination~\cite{chen2020auto}. These methods typically focus on optimization of critical optical elements of the system, requiring either direct access to the optical system~\cite{yang2022subwavelength}  or robust, differentiable simulations during optimization. Due to the complexity and operating costs of modern telescopes, modification to the underlying system or real-time access for network training would be infeasible. Instead, we focus on optimization of optical elements available to us in such a way that would not affect its operation, allowing us to side step these limitations.
\vspace{-1em}
\section{Super Resolving Adaptive Optics}
\label{sec:srao}
\vspace{-0.3cm}
This work augments existing MISR methods by learning the optimal way to enhance each individual image captured over a single exposure.
Using the DM already present in a telescope's AO system, we inject optimized phase profiles, thereby enabling multi-image super-resolution.

In this section, we present an end-to-end approach to determine optimal phase profiles while accounting for AO hardware constraints and atmospheric effects on final images. We discuss our approach's assumptions (\cref{ssec:small_phase}), the framework for optimizing phase profiles (\cref{ssec:phase_shift}), compatibility with various image reconstruction methods for combining sub-exposures (\cref{ssec:sr_methods}), and the training details for obtaining the optimal phase profiles (\cref{ssec:sr_methods}).

\vspace{-1em}
\subsection{The ``Small Phase Shift'' Assumption}
\label{ssec:small_phase}
\vspace{-0.3cm}
To include the image formation model in our learning infrastructure we must make some reasonable assumptions about the process. Importantly, we assume that sub-pixel shifts negligibly impact the AO loop or science image. In our case this is reasonable due to the fact that we consider only under-sampled systems. If our phase shifts keep $\earthPSF$ within the range of its under-sampled area, the aggregated effect over the entire exposure will be negligible\footnote{If necessary, we can also impose constraints on the magnitude of our learned phase shifts to guarantee science performance}. The shifts define the DM ``flat'' (a standard feature of AO real-time controllers), \edit{and can be updated easily between exposures}. Therefore, the AO system will run \edit{entirely as usual}---unaware of the phase offsets applied by our method. 

\vspace{-1em}
\subsection{Phase Shifting and Optimization}
\label{ssec:phase_shift}
\vspace{-0.3cm}
Because our DM is designed to create highly complex shapes, we are not restricted to planar movements like traditional methods~\cite{fruchter2002drizzle}. 
\edit{We describe a given DM shape $\shiftPSF$ using either actuator positions or, as detailed below, a more general modal basis.}

\replaced[]{Specifically, one approach would be}{One common approach is} to use a combination of Zernike polynomial modes, $\zernPSF$. The Zernike modes are well suited for this problem because of their ability to represent rich phases (including planar shifts) with minimal coefficients and are well-suited for circular apertures, such as those of telescopes~\cite{lakshminarayanan2011zernike}. They \added{can} also provide some insight into \replaced{the nature of a given phase due to the relationship between Zernike modes and classical optics aberrations (e.g., focus, coma, and astigmatism~\cite{cheng2010visual})}{the types of phase that is optimized due to the relationship between Zernike modes and classical optics aberrations such as focus, coma, and astigmatism~\cite{cheng2010visual}}. Besides planar shifts (tip and tilt Zernike modes), there is no trivial or well known choice of Zernike modes for super-resolution available. We therefore optimize their values in conjunction with the MISR algorithm of choice to estimate the best possible choice of modes (i.e., an end-to-end optimization).

\deleted{Beyond Zernike modes, one extension we explore is to simply optimize the entire phase profile, $\shiftPSF$, during training instead of relying on any particular basis. Later we will compare all three of these phase representations for comparison against baseline algorithms.}

We can now update our image formation model from a single exposure $\capturedImage_i$ of time $T$ to include $N$ exposures of time $T/N$ with individually optimizable phases $\shiftPSF_{i=1\dots N}$
\added{
\protect \begin{equation}
    \capturedImage_i = \abssq{\objectImage} \defconvolve \abssq{\systemPSF_i \defconvolve e^{\edit{j}\shiftPSF_i}} + \imageNoise.
    \label{eq:discrete_modal}
\end{equation}}
This same model can be equally applied to a simple shifting or Zernike model by replacing $\shiftPSF$ with $\zernPSF$ and choosing an appropriate number of Zernike coefficients.

Now, let $\funcUpsample$ be a learned MISR function with parameters $\theta$, and $\funcDownsample$ be a downsampling method which reduces an image's spatial dimensions by a fixed factor. Our end-to-end optical optimization problem is then,
\added{
\protect \begin{equation}
    \min_{\{\funcParamters,\,\shiftPSF\}} \Big|\Big| \funcUpsample[ \funcDownsample[\abssq{\objectImage} \defconvolve \abssq{\systemPSF_{1\dots N} \defconvolve e^{\edit{j}\shiftPSF_{1\dots N}}} + \imageNoise]] - \objectImage\Big|\Big|_1,
    \label{eq:loss_function}
\end{equation}}
where $\objectImage$ is an image randomly sampled from a set of high-resolution images, \replaced[comment={\JXeightR: \textit{Were PSFs for superresolution [...] a random sampling?}}]{$\systemPSF_i$ is randomly chosen}{$\defphase{S}_i$} from a set of pre-computed AO telescope PSFs, $\shiftPSF_i$ are our trainable phase profiles, and $||\cdot||_1$ denotes the $\ell_1$ norm.
\added[comment={\JXeightR: \textit{What are the methods in Table 1 and 2?}}]{As described earlier, $\shiftPSF$ can be represented in many ways. To explore its effects, we compare the most common methods in our experiments. These include:}
\begin{itemize}
    \item \textbf{No Learned Phase:} The base case for CNN methods. The same number of exposures are given to the CNN during training---and still affected by $\systemPSF_{\edit{i}}$---but without optimizing any per-exposure $\shiftPSF_{\edit{i}}$.
    \item \textbf{Tip/Tilt:} A common MISR technique where the image is translated by a sub-pixel shift~\cite{farsiu2003robust}.
    \edit{The \textit{Classic} case uses a sequence of exposures whose translational shifts form a regular sub-pixel raster scan. The \textit{Optimized} variant initializes with the same shifts as \textit{Classic} and then optimizes the values of those shifts.}
    \item \textbf{Zernike Modes:} Each phase is represented using the Zernike basis and its coefficients are optimized.
    \item \textbf{Non-Modal:} Instead of a basis function, the DM shape is directly optimized. This requires precise knowledge of its influence functions and calibration with respect to the WFS~\cite{berdeu2023analytical}. While simple in simulation, this proves to be difficult to use in practice.
\end{itemize}

\vspace{-1em}
\subsection{Super-Resolution Reconstruction Methods}
\label{ssec:sr_methods}
\vspace{-0.3cm}
Although our method could work equally well with any super-resolution reconstruction method, for the purposes of this paper we focus on only two:

First, we consider a linear-reconstruction method similar to Drizzle~\cite{fruchter2002drizzle}, also known as the shift-and-add method~\cite{farsiu2003robust}. This may be of interest in science cases where such methods are already widely used, albeit for systems without adaptive optics and with un-optimized phases.

Second, we consider a state of the art deep learning-based \edit{SR} algorithm. Specifically, we based our model on the widely known EDSR~\cite{lim2017enhanced} single-image \edit{SR} CNN. Our network was modified to incorporate the forward image formation model (\cref{eq:image_formation_model}) as well as to accept any number of input channels, representing the phase-shifted exposures. The EDSR model was chosen for its speed, widespread adoption, and its avoidance of adversarial training (i.e., it does not rely on a GAN~\cite{creswell2018generative}) which can complicate training and introduce priors that may not be appropriate for scientific purposes\footnote{Our outlined procedure can be easily adapted to any SR method including diffusion~\cite{li2022srdiff} or transformer~\cite{lu2022transformer} models, or traditional convex optimization methods~\cite{chan2016plug}.}.

\vspace{-1em}
\subsection{Phase and Network Training Details}
\label{ssec:traiing}
\vspace{-0.3cm}
In addition to a training set of high-resolution images, our phase and upsampling parameters require a collection of system PSFs under the influence of the AO system and the expected turbulence. These can be generated by simulations of the system or collected from telescope telemetry. 

At training time, for a model expecting $N$ exposures, we provide the network a random batch of high-resolution images as well as a random sample of $N$ system PSFs. The low-resolution images are generated for each PSF, as detailed in \cref{eq:discrete_modal}, using a trainable phase profile for each sub-exposure. We now have low-resolution images affected by the system PSF in addition to our optimized phases. We also have the ground-truth high-resolution image, which can be used to compute the loss function in \cref{eq:loss_function}.

Results shown in this work were trained on a combination of the DIV2K~\cite{agustsson2017ntire} and Microsoft ImagePairs~\cite{joze2020imagepairs} datasets. Both CNN and linear reconstructors were \edit{implemented and trained} with \texttt{pyTorch}~\cite{paszke2019pytorch} following the learning procedure described in the EDSR paper~\cite{lim2017enhanced}.
\added{Unless stated elsewhere (e.g., \cref{sec:num_exposure_tests}), the following experiments are all performed with $N=4$ exposures. While our method can be used with any number of exposures, this value was chosen to easily compare to the classic ``tip/tilt'' methods when solving for a $2\times$ upscaling factor. Restricting the number of exposures to four highlights the performance differences between linear and deep learning-based reconstruction methods, as discussed in ~\cref{sec:simulated_results,sec:experimental_results,ssec:phase_analysis}.}
\begin{figure*}[!ht]
    \centering
    \includegraphics[width=0.98\textwidth]{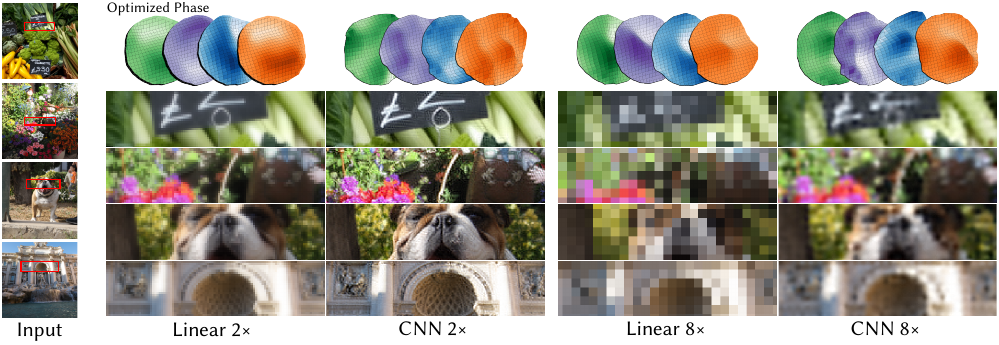}
    \vspace{-.4em}
    \caption[Simulated Super-Resolution Results]{\textit{Simulated SR results.} Comparisons of reconstructions with super resolution factors of 2$\edit{\times}$ and 8$\times$ (inset regions marked in red). Images were generated with $N=4$ exposures and phases were optimized for each method and \edit{SR factor}. For a full view of the results, please refer to the supplemental material~\cite{this_supplemental}.}
    \label{fig:sim_results_magnification}
\end{figure*}

\vspace{-1em}
\section{Simulated Results}
\label{sec:simulated_results}
\vspace{-0.3cm}
\edit{We first evaluate our method entirely in simulation}. 
We train SR parameters and phase profiles solely on synthetic PSFs generated by the discretized image‑formation model in~\cref{eq:discrete_modal}. 
After training, we run a new AO simulation, setting the optimized phase profiles as the deformable mirror’s ``flat'' position.
This imposes the optimized phase on the system, replicating our convolution in the simulated hardware.
The PSF generated from this new simulation is then used as the PSF for evaluating test images. 

\vspace{-1em}
\subsection{AO Simulation Environment}
\vspace{-0.3cm}
To create a realistic AO simulation we use the OOMAO~\cite{conan2014object} toolbox. We implemented a realistic AO simulation with multiple layers of atmosphere to generate realistic PSFs for training. Our simulated AO system was set up to match our experimental hardware (detailed in~\cref{sec:experimental_results}) with an 8~m diameter telescope and 397-actuator deformable mirror operating at 1000~Hz. 
Each PSF was generated by performing a full AO simulation with an evolving atmosphere over a 2-second exposure. The simulated $\earthPSF$ was therefore imaged on a simulated science camera under the effects of AO correction and realistic sensor noise. To cover a wide range of conditions, the wind direction, wind speed, and guide star magnitude were randomized for each simulation.

\begin{figure}

    \centering
    \includegraphics[width=0.47\textwidth]{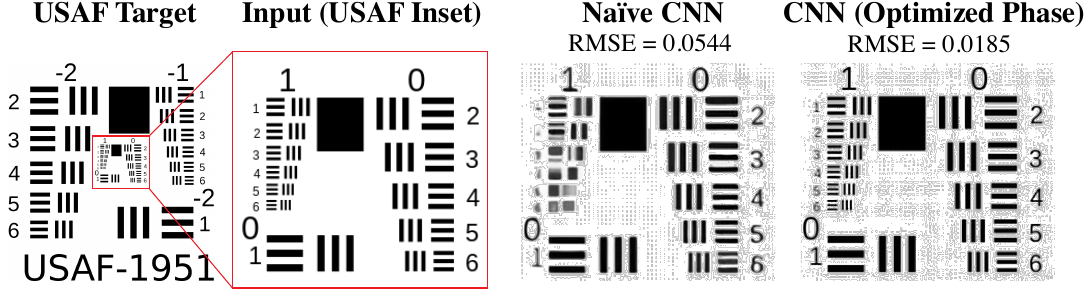}
    \vspace{-0.2em}
    \caption[Simulated Super Resolution Results]{\textit{Improved SR with jointly-optimized phase and CNN on simulated USAF resolution target.} End-to-end training of both phase and CNN yields superior resolution improvements compared to a naïve approach which does not optimize phases. The input to both CNNs is the inset region of the USAF target, marked in red. Qualitatively, by jointly optimizing the phase and the CNN, we can resolve smaller features and, quantitatively, achieve lower RMSE (displayed above each image).}
    \label{fig:usaf_chart}
\end{figure}

\noindent\textbf{Effect of Joint Optimization:}
We investigate how reconstruction results are affected by jointly optimizing phase profiles and the CNN reconstruction method. 
As shown in~\cref{fig:usaf_chart}, the jointly-optimized approach yields better resolution enhancement than a naïve approach that trains the CNN without phase optimization,
both qualitatively and quantitatively.

\noindent\textbf{Effect of \edit{Super-Resolution} Factor:}
We compare different reconstruction methods and phase representations for \edit{the} case where $N=4$ exposures. We compare our different phase representations against a baseline CNN case where the input to the neural network are four fixed unshifted PSFs with no optimizable phase and simple bilinear interpolation. Experimental results are shown in~\cref{table:sim_results}. Sample phase profiles and output can be found in~\cref{fig:sim_results_magnification}.

\begin{table}
\centering
\renewcommand{\arraystretch}{0.9} 
\resizebox{0.4\textwidth}{!}{%
\begin{tabular}{@{}lllll@{}} 
& & \multicolumn{3}{c}{Scale Factor} \\ \cmidrule{3-5}
& Method                 & 2$\times$ & 4$\times$ & 8$\times$ \\ \midrule \midrule
\multirow{5}{*}{CNN} & No Learned Phase           &  33.72  &  27.98  &  24.04  \\
 & Tip/Tilt (Classic)          &  39.56  &  33.46  &  25.47  \\
 & Tip/Tilt (Optimized)          &  39.67  &  33.88  &  26.32  \\
 & Zernike Modes      &  41.51  &  34.09  &  27.74  \\
 & Non-Modal          &  \textbf{43.13}  &  \textbf{34.24}  &  \textbf{27.79}  \\ \midrule
\multirow{4}{*}{Linear} & Tip/Tilt (Classic)       &  28.56  &  26.04  & 23.32  \\
 & Tip/Tilt (Optimized)       &  29.11  &  26.12  &  23.26  \\
 & Zernike Modes   &  30.84  &  26.17  &  23.36  \\
 & Non-Modal       &  31.08  &  26.67  &  23.33  \\ \bottomrule
 & & & & \\
\end{tabular}%
\label{table:sim_results}
}
\renewcommand{\arraystretch}{0.8} 
\resizebox{0.4\textwidth}{!}{
\begin{tabular}{@{}llllll@{}} 
& & \multicolumn{4}{c}{Number of Exposures} \\ \cmidrule{3-6}
& Method                 & 2 & 4 & 6 & 8 \\ \midrule\midrule
\multirow{3}{*}{CNN} & Tip/Tilt (Classic)          &  30.77  &  33.89  &  36.03  &  36.31  \\
 & Zernike Modes      &  31.13  &  34.11  &  \textbf{36.29}  &  37.44  \\ 
 & Non-Modal      &  \textbf{31.17}  &  \textbf{34.24} &  36.24  &  \textbf{37.48}  \\ \midrule
\multirow{3}{*}{Linear} & Tip/Tilt        &  26.11  &  26.12  &  26.12  &  26.11  \\
 & Zernike Modes   &  26.14  &  26.17  &  26.16  &  26.17  \\ 
 & Non-Modal   &  26.66  &  26.67  &  26.71  &  26.69  \\ \bottomrule
\end{tabular}}
\caption{
\textit{Effect of number of exposures and super-resolution factor on SR methods. }
Values presented in PSNR (dB) {$\uparrow$}. In general the more exposures, and the more expressive the phase basis, the better the results. Details on each method can be found in~\cref{ssec:phase_shift}.}
\label{table:sim_results_subexp}
\end{table}

\noindent\textbf{Effect of Number of Exposures:}
\label{sec:num_exposure_tests}
We consider the effect of number of exposures on each method, freezing the super-resolution factor. The results are shown in~\cref{table:sim_results_subexp}.

\vspace{-1em}
\section{Experimental Results}
\label{sec:experimental_results}
\vspace{-0.3cm}

\begin{figure}[ht]
  \centering
  \includegraphics[width=.35\textwidth]{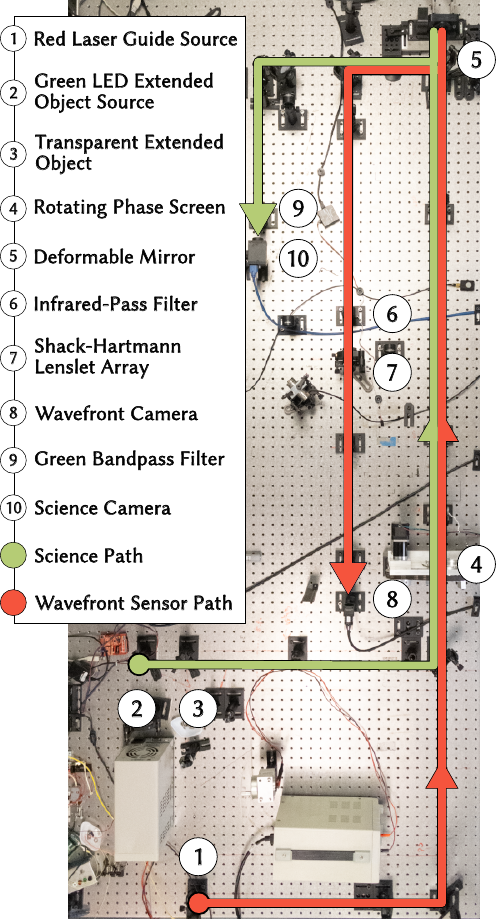}
    \vspace{-.5em}
  \caption[Experimental Optical Bench Layout]{\textit{Optical setup.} We present a top-down view of the optical bench where our experiments were conducted, labeling both the wavefront sensor path and the science (image) paths.}
  \label{fig:experimental_bench}
\end{figure}

To demonstrate our method's ability to work in realistic settings, we also implemented the experimental setup described in \cref{sec:simulated_results} on an optical table. Our setup, shown in~\cref{fig:experimental_bench}, consists of a 97-actuator ALPAO deformable mirror and a custom-built $25\times25$ sub-aperture Shack-Hartmann wavefront sensor. 
Our guide star was created with an 850~nm near-infrared laser whose pupil and beam size were matched to the DM's size (13.5~mm). Realistic atmospheric turbulence was induced by a rotating Lexitek\protect\footnote{Lexitek Inc. Phase Plates -- \url{https://lexitek.com}} phase screen driven by a stepping motor. The sensor's native pixel pitch was 5.86~$\mu$m and the diffraction-limited spot was \edit{22.95}$~\mu$m ($f$~=~400~mm, pupil size 13.5~mm, $\lambda$~=~635~nm). Note that even though our experimental hardware was not undersampled by design, we used 12$\times$ on-sensor binning to capture individual images.  This resulted in an effective pixel pitch of 70.3~$\mu$m.

The AO loop was controlled with the open source $\texttt{pyRTC}$ software\footnote{pyRTC on Github: \url{https://github.com/jacotay7/pyRTC}} which allows full control of the DM and science camera. This allowed us to define the optimized phase shifts \edit{to be} the reference mirror ``flat'', from which the residual wavefront was measured by the AO WFS. Each image was then captured as usual, changing the DM's ``flat'' between exposures as dictated by the optimized phases, and then combining these images to restore our final SR image.
\vspace{-1.em}
\subsection{Results}
\label{ssec:exp_results}
\vspace{-0.3cm}
\begin{figure}[ht]
     \centering
     \includegraphics[width=0.45\textwidth]{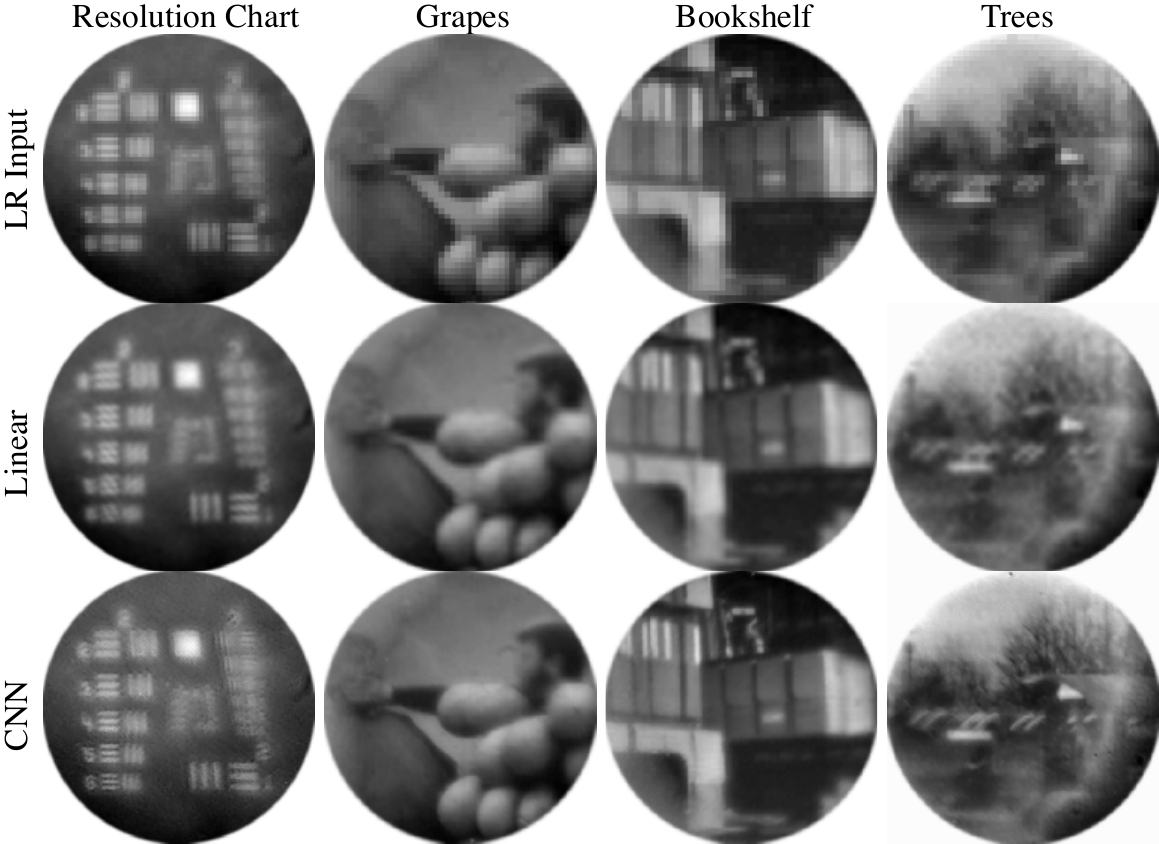}
    \vspace{-.25em}
        \caption[Experimental Super Resolution Results on USAF Resolution Chart]{\textit{Experimental results.} We reconstruct $8\times$ upsampled images of the USFA resolution target and three natural images (Grapes, Bookshelf, Trees) from low resolution images (LR Input) captured with the optical setup depicted in ~\cref{fig:experimental_bench}.  We present results using both linear and CNN method, each using their own optimized phase profiles. Both result in higher resolutions than the original input, as seen quantitatively on the USFA chart.}
        \label{fig:exp_results_natural}
\end{figure}
A green LED  (centered at 530~nm) was used to project photographic slides on the optical bench detailed above. To demonstrate \edit{our method's} resolving power, we imaged a $1"$ US Air Force resolution chart (Thorlabs Positive 1951 USAF Test Target Groups 2-7, Ø1"). \edit{Results are shown} for both simulation (\cref{fig:usaf_chart}) and experiment (\cref{fig:exp_results_natural}). 
We also show our method’s performance on more general scenes using a variety of real-world photos in~\cref{fig:exp_results_natural}.
\edit{Because our AO system was designed specifically for sensing in infrared wavelengths, the quality of our images (captured at 635 nm) is limited by chromatic aberration and vignetting.}
As such, results in this paper are cropped and masked for clarity; raw images are provided in the supplemental materials~\cite{this_supplemental}.
\vspace{-1em}
\section{Discussion}
\label{ssec:mtf_analysis}
\begin{figure}[!ht]
    \centering
    \includegraphics[width=\linewidth]{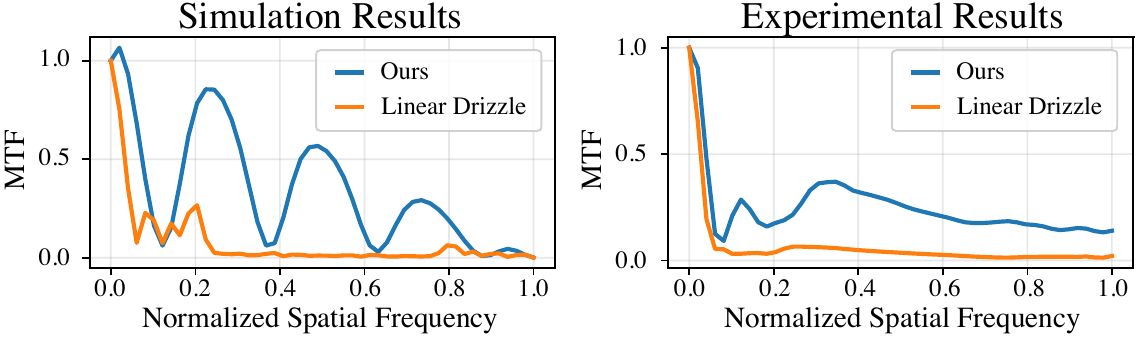}
    \vspace{-1.5em}
    \caption[Comparison of simulated and experimental MTFs]
    {
    \textit{Comparison of simulated and experimental MTFs.} 
    }
    \label{fig:mtf_analysis}
\end{figure}

\vspace{-0.3cm}
\textbf{MTF Analysis:} The modulation transfer function (MTF) is a measure of the achievable contrast for an optical system over a range of spatial frequencies~\cite{nasse2008read}. An ideal lens will linearly decrease from an MTF of 1 down to 0 as the spatial frequency increases and, in general, the closer a measured MTF behaves to this the better its resolving power. The MTF plots in \cref{fig:mtf_analysis} correspond to the USAF simulation of \cref{fig:usaf_chart} and the real experiment of \cref{fig:exp_results_natural} computed using a knife-edge method~\cite{li2016measurement}. Both show an increase of MTF across all spatial resolutions and thus demonstrate our method's superiority (i.e., higher preserved contrast). 

\begin{figure}[!ht]
    \centering
    \includegraphics[width=.475\textwidth]{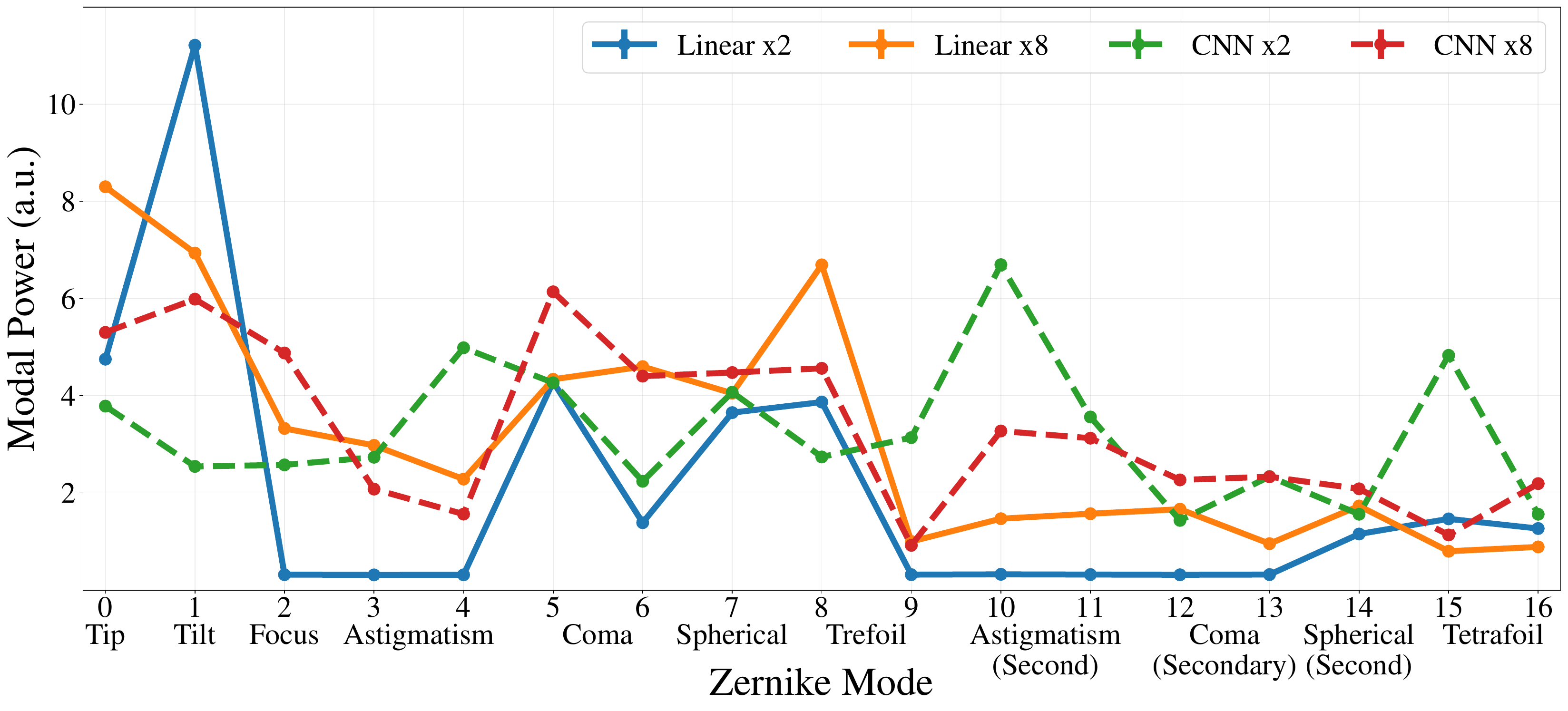}
    \vspace{-2em}
    \caption[Modal powers used to induce SR]{\textit{Comparison of modal powers used to induce SR.} We show modal power in the Zernike basis to explore the optimized phases (corresponding to the phases shown in \cref{fig:sim_results_magnification}). The linear method uses low-order, symmetric modes while the CNN spreads across all modes. Comparing $8\times$ vs. $2\times$ magnifications, they take opposite approaches: the CNN prefers low modes at higher resolutions, while the linear uses lower modes at low resolutions.}
    \label{fig:phase_analysis_power}
\end{figure}

\label{ssec:phase_analysis}
\noindent\textbf{Phase Analysis:} One advantage of representing phase profiles using Zernike modes is the ability to infer exactly which optical aberrations are most important to our reconstruction algorithms.
\edit{In \cref{fig:phase_analysis_power}, we plot the total absolute power of the first nine basis modes identified by our method for both the CNN and the linear reconstructor, using a fixed number of exposures while varying the magnification factor.}
\edit{The optimized phases reveal insights.}
First, total modal power increases with the magnification order. Intuitively, this makes sense as more movement within a pixel is needed to sample higher resolutions. Second, CNNs apply roughly the same power across all modes for a given magnification while the linear method has more variance. \edit{Perhaps} this is because linear solvers \edit{cannot} leverage higher-frequency spatial information in the image as well as CNNs can.

\noindent\textbf{Limitations \& Future Work:} Currently, our work is limited to telescopes with a single-conjugate adaptive optics (SCAO) system, which treats the atmosphere as a single column. Many of today's observatories instead use multiple WFS and DMs. This leads to more accurate corrections across a wider FoV and can even be tuned to different atmospheric layers. Such systems do not preclude the use of our method, since our method can be extended to multiple DMs and potentially enhanced by their extra degrees of freedom.

Other directions for future work include real-time viewing of video from the science camera by cycling through each phase multiple times to create individually super-resolved frames. Once extended to multiple DMs, additional multiplexing could also be achieved by varying each DM at different rates.

Most importantly, now that we have demonstrated our methodology with an optical bench prototype, our current priority is to show its effectiveness on sky with a science class telescope and AO system.

\noindent\textbf{Conclusion:}
\edit{This work shows that science images can encode sufficient MISR information using the AO systems already present in state-of-the-art telescope observatories.}
\edit{We achieve resolution gains without affecting the scientific data in both simulations and optical-bench experiments.}
With this method, we aim to unlock new scientific capabilities at existing observatories without altering the optical system.

\vspace{-1em}
\section*{\edit{Acknowledgements}}
\vspace{-0.3cm}
\edit{RS acknowledges NVIDIA for GPU hardware support. EYHL acknowledges the support of NSERC under CGSD. SS acknowledges the support of the Canada Foundation for Innovation and the Ontario Research Fund. KNK acknowledges the support of NSERC under the RGPIN and RTI programs.}
{
    \small
    \vspace*{-1em}
    \bibliographystyle{ieeenat_fullname}
    \bibliography{main}
}

\clearpage
\setcounter{page}{1}
\maketitlesupplementary

\section{The Shack-Hartmann Wavefront Sensor}

\begin{figure*}
    \centering
    \begin{subfigure}[T]{.595\textwidth}
        \centering
        \includegraphics[width=\textwidth]{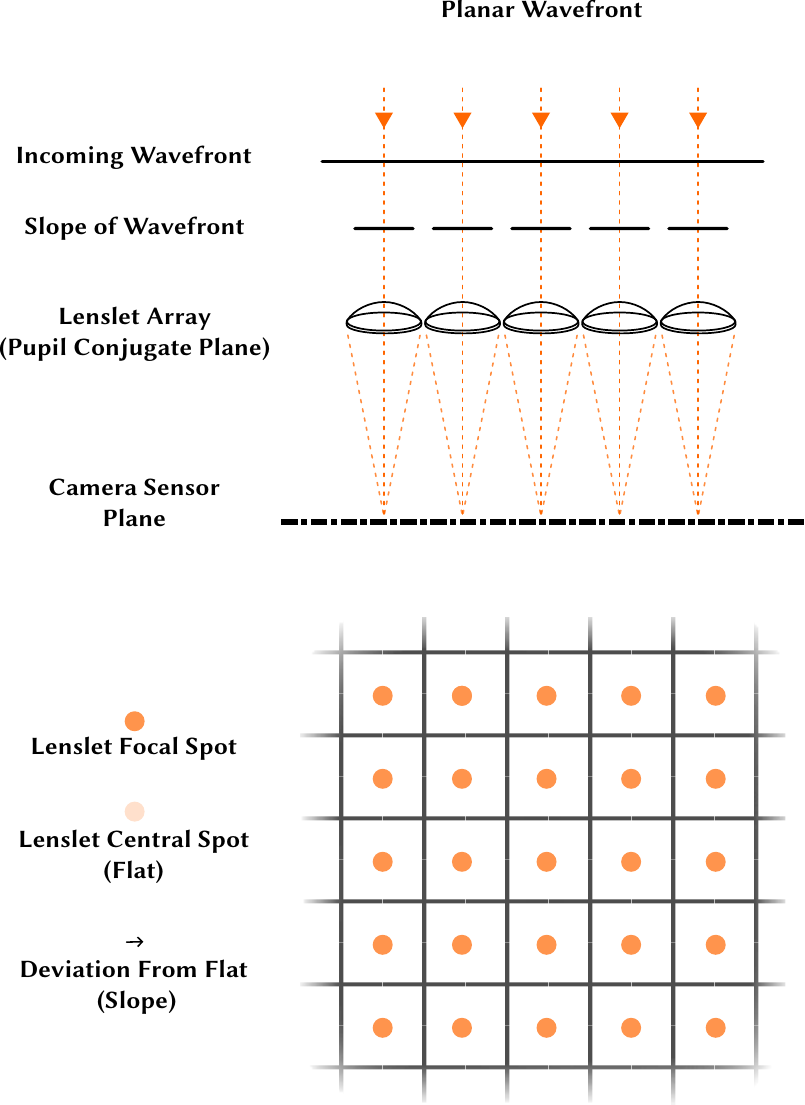}
        \vspace{.5em}
        \caption{} 
    \end{subfigure}
    \begin{subfigure}[T]{.39\textwidth}
        \centering
        \includegraphics[width=\textwidth]{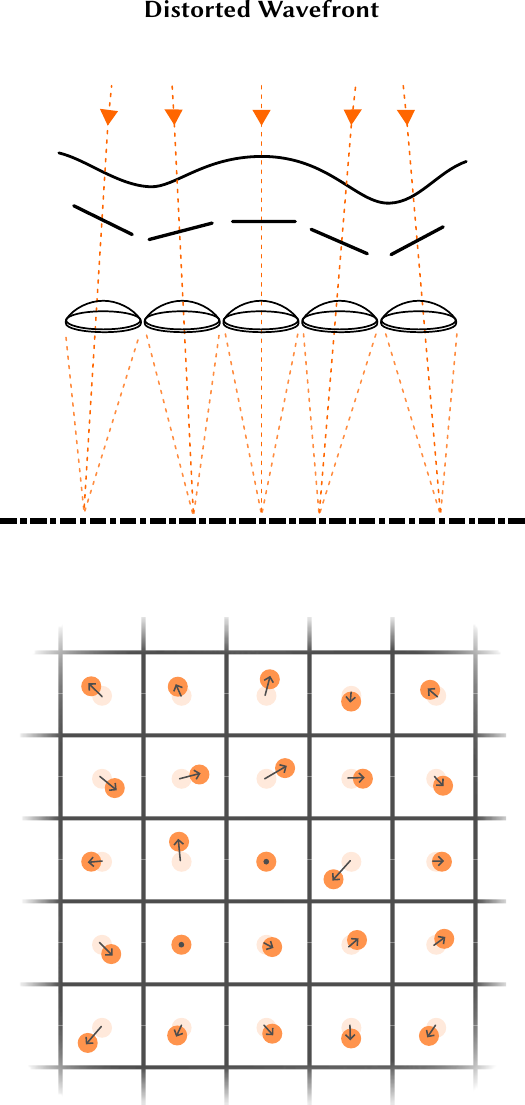}
        \vspace{.5em}
        \caption{} 
    \end{subfigure}

    \caption[The Shack-Hartmann Wavefront Sensor]{\textit{Basic working principles of the Shack-Hartmann Wavefront Sensor}. If a perfectly flat wavefront reaches the SHWFS (a), each lenslet will focus the guide star as a point in the middle of its corresponding pixel grid. Any deviation to the wavefront (b) will result in an $x$ and/or $y$ shift in the sensor plane. These deviations directly correspond to the $x$ and $y$ ``slope'' (first derivative) of the wavefront.}
    \label{fig:incoming-slope}
\end{figure*}

\begin{figure*}[ht]
  \centering
  \includegraphics[width=0.99\textwidth]{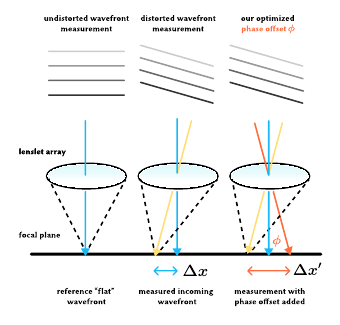}

  \caption[Illustration of how a new flat is induced on a Shack-Hartmann wavefront sensor]{\textit{The Shack-Hartmann Wavefront Sensor (SHWFS)}: Wavefront distortions can be measured by focusing incoming light to a point and measuring its $(x,\,y)$ position with respect to a reference ``flat'' position. Using a grid of lenslets, the SHWFS can spatially sample the wavefront across the entire image sensor. This signal is sent to the control system which determines the ideal mirror positions to remove the deviations. Our method changes the reference positions, applying a small offset to each lenslet, inducing an optimized phase shift on the mirror.}
  \label{fig:ao_shwfs_schematic}
\end{figure*}

\blfootnote{Corresponding author: \url{robin@cs.toronto.edu}}
\blfootnote{Project website: \url{www.cs.toronto.edu/~robin/aosr}}
Here we include additional details on the operation of the Shack-Hartmann Wavefront Sensor (SHWFS) in \cref{fig:incoming-slope} and how a new flat is applied to the AO system in \cref{fig:ao_shwfs_schematic}.

\begin{figure*}
     \centering
      \includegraphics[width=0.95\textwidth]{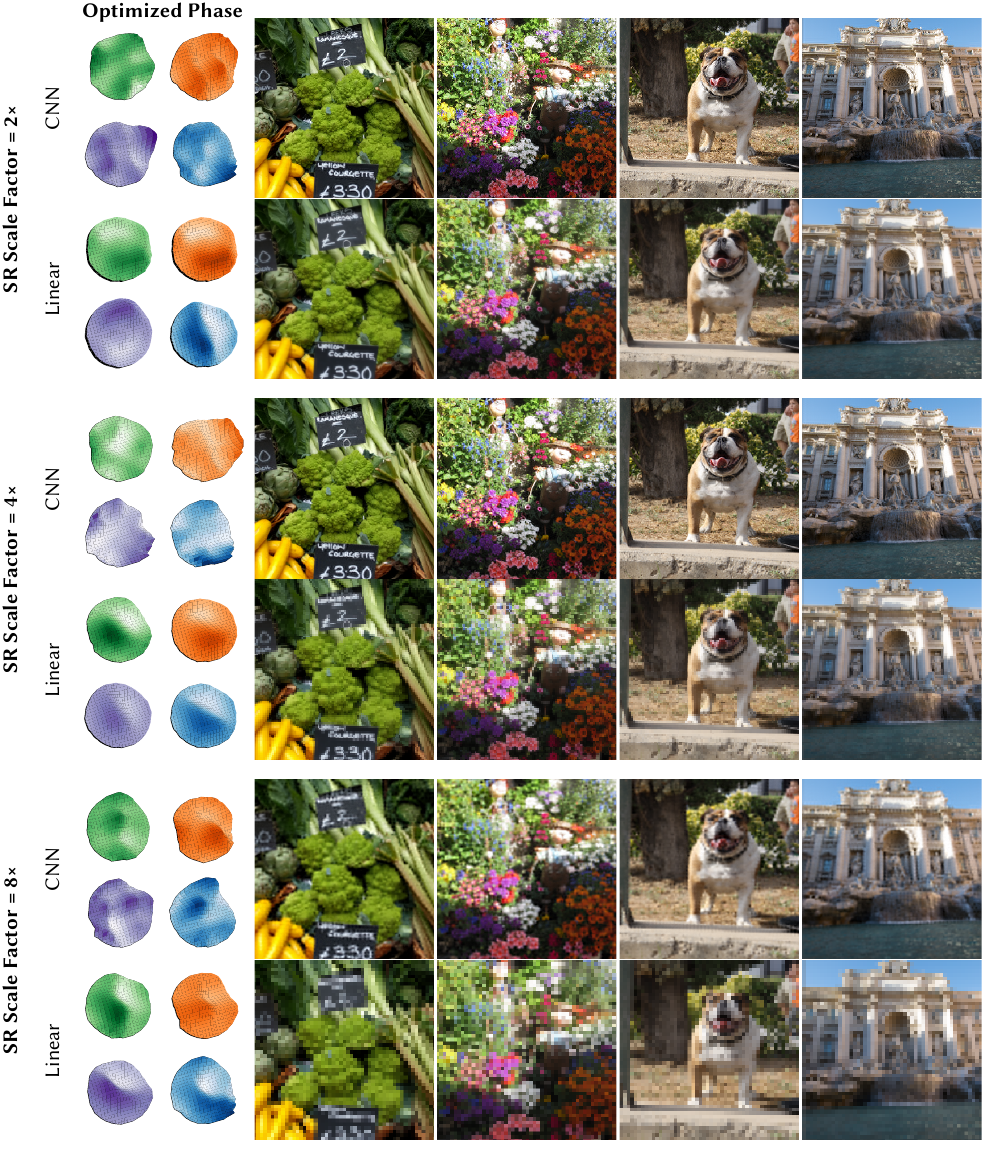}
        \caption[Simulated Super-Resolution results]{\textit{Simulated Scaled Results with $N=4$ Exposures from the PIRM Dataset} }
        \label{fig:sim_results_supplemental}
\end{figure*}

\section{Simulation Settings}
\label{sec:sim-settings}

\begin{table}[!h]
\renewcommand{\arraystretch}{0.9} 
\resizebox{0.5\textwidth}{!}{
\setlength{\tabcolsep}{16pt}
\centering
\begin{tabular}{ccc} \toprule
\multicolumn{2}{c}{Simulation Parameters}                           & Values                                                        \\ \midrule
\multirow{9}{*}{Telescope}      \ 
                                & Diameter                          & 8 m                                                           \\
                                & Sampling Frequency                & 800 Hz                                                        \\
                                & WFS Order                         & $16\times16$                                                  \\
                                & WFS Readout Noise                 & $\approx0 e^{-}$                                              \\
                                & DM Order                          & $17\times17$                                                  \\
                                & NGS Band                          & R                                                             \\
                                & NGS Magnitude                     & $\mathcal{U}(8, 16)$                                          \\ 
                                & POL Gain                          & $0.35$                                                        \\ \midrule
\multirow{12}{*}{\vtop{\hbox{\strut Three Layer}\hbox{\strut Atmosphere}}} \ 
                                & $r_0$                             & $\mathcal{N}(0.15, 0.02)$ cm                                  \\ 
                                & Layers                            & $3$                                                           \\ \cmidrule(r){2-3}
                                & \multirow{3}{*}{Altitudes}        & $0$ km                                                        \\
                                &                                   & $4$ km                                                        \\
                                &                                   & $10$ km                                                       \\ \cmidrule(r){2-3}
                                & \multirow{3}{*}{Fractional $r_0$} & $0.70$                                                        \\
                                &                                   & $0.25$                                                        \\
                                &                                   & $0.05$                                                        \\ \cmidrule(r){2-3}
                                & \multirow{3}{*}{Wind Speeds}      & $\mathcal{N}\left(5, 2.5\right)$ km/s                         \\
                                &                                   & $\mathcal{N}\left(10, 5\right)$ km/s                          \\
                                &                                   & $\mathcal{N}\left(25, 10\right)$ km/s                         \\ \cmidrule(r){2-3}
                                & Wind Directions                   & $\mathcal{U}\left[0, 2\pi\right)$ rad                         \\ \midrule
\multirow{2}{*}{Science Camera}        & Science Camera Band               & K                                                             \\
                                & \added[comment={\JXeightR: \textit{Zero noise in the supp table is a bit suspicious.}}]{Science Camera Noise}                 & $1\%$ Max Value                                              \\
\bottomrule
\end{tabular}}
\caption{\textit{Simulation parameters used for generating point spread functions and the training of our reconstruction methods.}\label{tab:simulationsettings}} 

\end{table}


\section{Experimental Calibration}
\label{sec:calibration}

To calibrate the modal power needed to match simulation and experiment, we simply induce a change to the flat with a single mode applied with different amounts of power. We can then compare the output from our forward model with varying amounts of the same mode and find the scalar value that best matches. This process can be repeated for any number of modes to ensure linearity across each mode and their power. As previously mentioned, the PWFS response is linear only about some working point and so this may be more important when using one compared to the SHWFS used in these experiments.

We also use test-time augmentation in order to re-normalize the network and match the new experimental data distribution. We employed a modified version of the popular TENT~\cite{wang2020tent} algorithm where only the batch-normalization scalers are updated using cropped areas of the input data and their bilinear-upsampled values as input/training pairs.

\section{Additional Experimental Results}
\label{sec:uncropped}

\begin{figure*}[ht]
  \centering
  \includegraphics[width=0.99\textwidth]{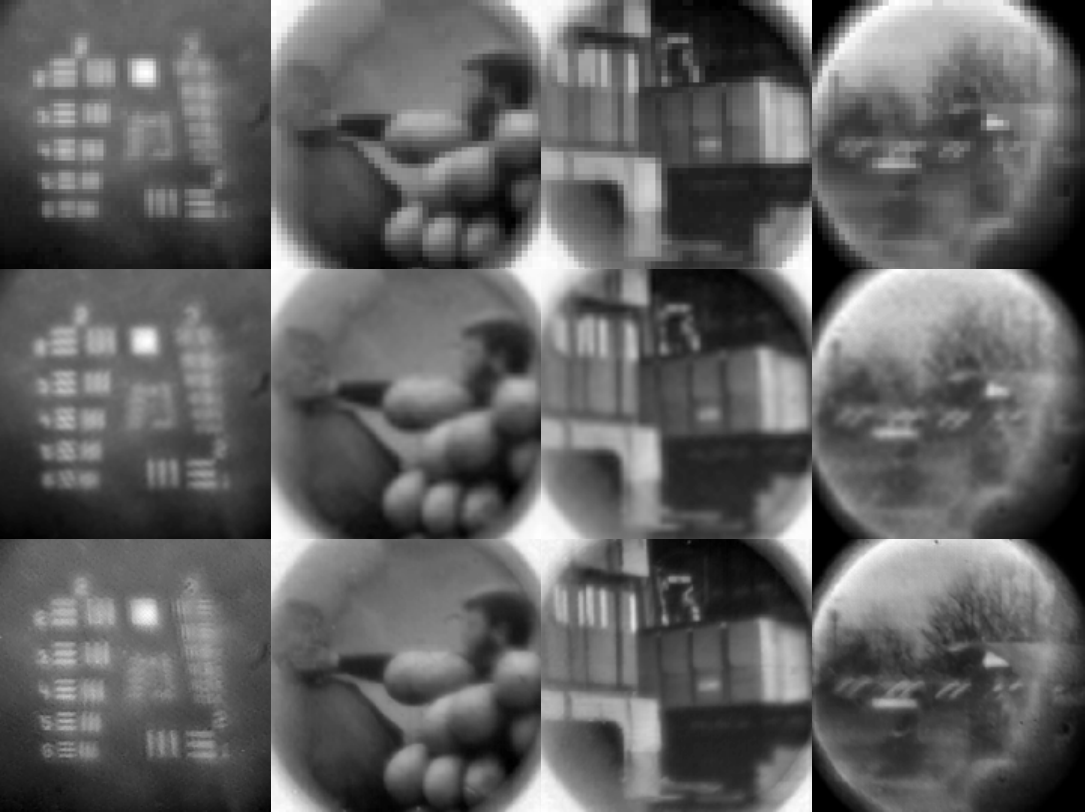}
    \caption{\textit{Uncropped Experimental Results.}}
    \label{fig:uncropped-experimental-results}
\end{figure*}

Here we include additional simulated results in \cref{fig:sim_results_supplemental} as well as the larger, uncropped experimental results from \cref{ssec:exp_results} in \cref{fig:uncropped-experimental-results}. 

\end{document}